\documentclass[]{interact}

\usepackage{epstopdf}% To incorporate .eps illustrations using PDFLaTeX, etc.
\usepackage[caption=false]{subfig}% Support for small, `sub' figures and tables
\usepackage{longtable}
\usepackage{siunitx}
\usepackage{booktabs,tabularx,ragged2e,dcolumn,caption} 
\usepackage[flushleft]{threeparttable}
\usepackage{amsmath}
\usepackage{mathpazo}
\usepackage{mathtools}
\usepackage[ruled,vlined]{algorithm2e}
\usepackage{array}
\newcolumntype{L}[1]{>{\raggedright\let\newline\\\arraybackslash\hspace{0pt}}m{#1}}
\newcolumntype{C}[1]{>{\centering\let\newline\\\arraybackslash\hspace{0pt}}m{#1}}
\newcolumntype{R}[1]{>{\raggedleft\let\newline\\\arraybackslash\hspace{0pt}}m{#1}}
\newcolumntype{d}[1]{D{.}{,}{#1}}

\usepackage[numbers,sort&compress]{natbib}% Citation support using natbib.sty
\bibpunct[, ]{[}{]}{,}{n}{,}{,}% Citation support using natbib.sty
\renewrobustcmd{\bfseries}{\fontseries{b}\selectfont}

\newcommand{\matr}[1]{\bm{#1}}

\theoremstyle{plain}% Theorem-like structures provided by amsthm.sty

\theoremstyle{definition}

\theoremstyle{remark}

\begin{document}

\articletype{ORIGINAL RESEARCH}% Specify the article type or omit as appropriate

\title{Dynamic Learning and Productivity for Data Analysts: A Bayesian Hidden Markov Model Perspective}

\author{
\name{yueyin.io@gmail.com\textsuperscript{a}\thanks{CONTACT Yue Yin. Email: yueyin.io@gmail.com}\thanks{}}
% \affil{\textsuperscript{a}Kellogg School of Management, Northwestern University, Evanston, USA}
}

\maketitle

\begin{abstract}
Data analysts are essential in organizations, transforming raw data into insights that drive decision-making and strategy. This study explores how analysts’ productivity evolves on a collaborative platform, focusing on two key learning activities: writing queries and viewing peer queries. While traditional research often assumes static models, where performance improves steadily with cumulative learning, such models fail to capture the dynamic nature of real-world learning. To address this, we propose a Hidden Markov Model (HMM) that tracks how analysts transition between distinct learning states based on their participation in these activities.

Using an industry dataset with 2,001 analysts and 79,797 queries, this study identifies three learning states: novice, intermediate, and advanced. Productivity increases as analysts advance to higher states, reflecting the cumulative benefits of learning. Writing queries benefits analysts across all states, with the largest gains observed for novices. Viewing peer queries supports novices but may hinder analysts in higher states due to cognitive overload or inefficiencies. Transitions between states are also uneven, with progression from intermediate to advanced being particularly challenging. This study advances understanding of into dynamic learning behavior of knowledge worker and offers practical implications for designing systems, optimizing training, enabling personalized learning, and fostering effective knowledge sharing.

\end{abstract}

\begin{keywords}
Dynamic Learning Models; Productivity; Learning-by-Doing; Knowledge Sharing; Hidden Markov Model; Bayesian Hierarchical Modeling
\end{keywords}

% \begin{hypo}
% Past experience of viewing queries written by colleagues has a negative effect on data analyst productivity. 
% \end{hypo}
% \end{subhypo}

% Figure \ref{fig:Hist_1} shows that the empirical distribution of all \emph{First Completion Time}s in our study is right-skewed.

% \begin{figure}[ht]
% \centering
% \includegraphics[width=\textwidth]{Hist_1.pdf}
% \caption{The empirical distribution of First Completion Time is right-skewed.}
% \label{fig:Hist_1}
% \end{figure}

\section{Introduction}
\subsection{Context}
Data analysts are vital to modern organizations, transforming raw data into actionable insights that inform decision-making. Their contributions enable businesses to identify trends, understand customer preferences, and improve operational efficiency. For example, retail companies rely on analysts to determine top-selling products during holiday seasons or forecast future consumer behavior. Similarly, social media platforms depend on analysts to identify the types of content that generate the highest user engagement. These insights are critical for shaping strategies that help organizations remain competitive in fast-paced markets.

A key aspect of a data analyst’s work is writing and executing SQL (Structured Query Language) queries. These queries allow analysts to extract and organize data from large databases, perform calculations, and generate reports. For instance, an analyst might write a query to identify the ten most searched items on an e-commerce website during a given month or calculate the average time users spend browsing certain product categories. The ability to create efficient queries is essential for addressing business questions quickly and accurately. The productivity of data analysts directly influences organizational performance. Faster and more accurate query writing leads to quicker insights, which are particularly valuable for time-sensitive decisions such as marketing adjustments during a product launch or supply chain realignments.

Developing proficiency in query writing is not a straightforward process. It is a learned skill that requires mastering technical tools, understanding database structures, and refining problem-solving strategies. Organizational learning theory defines learning as a change in knowledge that occurs through experience at the individual, group, or organizational level \cite{fiol1985organizational, crossan1999organizational}. Much of the early research on learning focused on manufacturing and service industries \cite{benkard1999learning, darr1995acquisition}, but more recent studies emphasize knowledge-intensive fields such as IT consulting and software development \cite{fong2007learning, kim2012learning, kang2009learning}.

Our empirical setting is a collaborative platform where data analysts access large-scale data, write and execute queries, and view queries created by their peers. This platform facilitates both independent problem-solving and collaborative knowledge sharing. For example, an analyst tasked with identifying e-commerce trends might review peer queries to understand alternative approaches, adapt existing solutions, or gain inspiration for tackling new challenges. These collaborative features foster a rich environment for both individual and collective learning, enhancing query-writing efficiency and accuracy.

This study focuses on two key modes of learning for data analysts: writing queries (direct experience) and viewing peer queries (indirect experience). Writing queries exemplifies "learning-by-doing," where individuals gain knowledge through hands-on experience. The concept of learning-by-doing has been extensively studied in individual learning curves. \cite{ebbinghaus2013memory} demonstrated how repetition improves memory retention, while Wright \cite{wright1936factors} observed cost reductions in airframe production through cumulative experience, formalizing the idea of the learning curve \cite{argote2012organizational}. More recent studies confirm the value of direct experience across industries. For example, software developers improve productivity through specialized tasks \cite{fong2007learning}, and IT consultants enhance problem resolution times through experience \cite{kim2012learning}. Evidence from healthcare and open-source software projects further underscores the benefits of cumulative experience \cite{kc2012accumulating, clark2013learning, singh2011hidden}.

Indirect experience, represented by viewing peer queries, offers complementary benefits. Analysts gain insights by observing how colleagues approach problems, adapt solutions, or solve complex tasks. Research on organizational learning highlights the value of shared knowledge and collaborative interactions \cite{levitt1988organizational, huber1991organizational}. The situative perspective emphasizes that learning is inherently social and interactive \cite{tyre1997situated}. Empirical evidence supports the importance of peer-based learning across contexts. For instance, repair technicians share practical knowledge through storytelling \cite{brown2000organizational}, and programmers leverage online platforms like Stack Overflow to acquire new skills and address technical challenges efficiently \cite{vasilescu2013stackoverflow}. However, the benefits of indirect experience are not universal. For example, \cite{waldinger2011peer} found no evidence of peer effects on researcher productivity in certain scientific fields.

Combining direct and indirect experience has been shown to enhance learning outcomes in diverse settings, from manufacturing to collaborative software development \cite{reagans2005individual, kc2013learning}. For data analysts, understanding the interplay between these two modes of learning is essential for designing effective training programs and fostering productive collaboration. This study contributes to this understanding by examining how these learning modes shape productivity within a collaborative platform.

\subsection{Research Gap and Motivation}
Most studies on learning assume static models where learning happens at a constant rate over time. These models also assume that an individual's ability, which affects performance, remains stable. However, these assumptions do not reflect the reality of learning processes. Individuals learn at different rates, and even the same person’s learning rate can change over time. As experience builds up, it contributes to "human capital," which can enhance an individual’s ability to learn and apply new knowledge in the future \cite{cohen1990absorptive}.

The concept of \emph{absorptive ability}, introduced by \cite{cohen1990absorptive}, emphasizes the importance of prior knowledge in acquiring and using new information. Studies in cognitive and behavioral science have shown that past experiences influence how individuals approach new learning tasks \cite{ellis1965transfer, lindsay2013human, bower1981theories}. Static models fail to capture these dynamics, leading to biased results when evaluating learning activities. For example, \cite{manski1993identification} pointed out the \emph{reflection problem}, where static models mistakenly attribute individual performance to aggregated group effects. This issue can misrepresent the actual drivers of learning and productivity. 

Dynamic models provide a better way to study learning behaviors over time. For instance, \cite{chan2014learning} examined how peer-based learning accumulates and changes the productivity of salespeople. Similarly, \cite{singh2011hidden} developed a stochastic model to study how developers in open-source projects improve their skills. These studies highlight the value of dynamic models, but there is limited research on how such models can be applied to individual learning in knowledge-intensive fields like data analytics. 

This study addresses this gap by proposing a Hidden Markov Model (HMM) to capture the dynamic learning processes of data analysts. The HMM framework models how analysts’ productivity and learning ability change over time as they engage in different learning activities. These activities include \emph{writing queries} (learning from direct experience) and \emph{viewing peer queries} (learning from indirect experience), which together build their knowledge and expertise. The model represents each data analyst as transitioning between a finite set of states, where each state reflects a specific level of productivity and learning ability. Transitions between states depend on participation in learning activities. Additionally, the model accounts for individual differences in learning through random effects, capturing unobserved factors like personal preferences or past experiences. To estimate the model, we use a hierarchical Bayesian approach with Markov Chain Monte Carlo (MCMC) techniques. This approach allows us to analyze variations across individuals and provides reliable estimates by combining prior knowledge with observed data. By applying this model to data on query-writing activities, we aim to reveal patterns in learning and productivity, offering new insights into dynamic learning processes. 

Previous research \cite{yin2018learning} investigated the effects of learning-by-doing and learning-by-viewing on data analyst productivity, with a focus on how peer characteristics, such as productivity and popularity, influenced the effectiveness of indirect peer learning. While insightful, this earlier work relied on a static framework that assumed constant learning effects over time. Building on the same empirical setting, this study adopts a dynamic perspective to examine how analysts transition between distinct learning states. By employing a HMM framework, it captures the evolution of learning behaviors and provides a deeper understanding of the relationship between productivity, learning activities, and state transitions. This approach addresses the limitations of static models and offers novel insights into the dynamics of learning processes in knowledge-intensive environments.

\subsection{Key Results and Contribution}
The key findings from our analysis are summarized below:
\begin{enumerate}
    \item \textbf{Fynamic Learning States and Productivity Patterns} The HMM identified three distinct learning states: novice, intermediate, and advanced. Analysts in higher states demonstrated greater intrinsic productivity, as reflected by reduced query completion times. The results confirm that learning is not static but evolves dynamically as analysts participate in different activities.
    \item \textbf{Differential Effects of Learning Activities} The effects of learning activities—\emph{writing queries} (learning by doing) and \emph{viewing peer queries} (learning by observing)—varied significantly across learning states. Writing queries positively influenced transitions to higher states for both novice and intermediate analysts, while viewing peer queries helped novices but hindered analysts in intermediate and advanced states. This highlights the importance of tailoring learning activities to the current state of the analyst.
    \item \textbf{Learning State Transition Dynamics} The model revealed that it is significantly harder for analysts to move from the intermediate to the advanced state than from novice to intermediate. This finding underscores the diminishing returns of learning activities at higher levels of expertise and suggests a need for tailored strategies to sustain learning momentum in advanced analysts.
\end{enumerate}

The contribution of this study is three-fold:
\begin{enumerate}
    \item This research advances the understanding of learning dynamics by introducing a dynamic framework that incorporates state transitions and heterogeneous effects of learning activities. Unlike static models, the proposed HMM captures the evolving nature of productivity and learning, shedding light on how prior knowledge influences future learning behaviors. This contributes to the broader literature on organizational learning and individual skill development.
    \item The study employs an innovative Hidden Markov Model with hierarchical Bayesian estimation, enabling the analysis of individual-level learning trajectories. By integrating random effects, the model accounts for unobserved heterogeneity, providing robust and personalized insights. The methodological approach offers a new way to study learning dynamics in knowledge-intensive tasks, which can be adapted to other domains.
    \item The findings provide actionable insights for designing adaptive systems and interfaces in collaborative platforms. By identifying learning states and tailoring activities accordingly, platforms can optimize both individual productivity and learning outcomes. For example, the results inform strategies for designing onboarding programs, personalized learning paths, and feedback mechanisms that support continuous development. These insights are particularly valuable for managers and system designers aiming to enhance productivity in knowledge-driven organizations.
\end{enumerate}

The remainder of this paper is structured as follows. Section 2 presents the materials and methods, detailing the empirical setting, data sources, and measurement approaches, including the dependent and related variables. This section also provides a comprehensive introduction to the HMM framework, covering its theoretical foundation, model development, and estimation strategy. Section 3 reports the key results, highlighting the findings on learning dynamics and productivity patterns. Finally, Section 4 discusses the implications of these findings, outlines limitations, and suggests directions for future research. The disclosure and data availability statements are shared in section 5 and section 6.

\section{Material and Methods}
\subsection{Empirical Setting}
This study investigates the behavior of data analysts engaging with a collaborative data platform widely adopted in industry for managing large-scale analytics workflows. The platform serves as an integrated environment for data analysts, providing functionalities for query composition, execution, and knowledge sharing. 
First, the platform acts as a centralized repository for metadata in analytics data warehouses, enabling seamless access to data from various services, such as relational databases (Teradata, MySQL, SQL Server, etc) and visualization tools like Tableau. Analysts with appropriate permissions can access datasets and integrate these tools into their workflows efficiently to produce comprehensible results.
Second, the platform supports a vibrant collaborative ecosystem beyond data access and analysis. A data analyst can share her knowledge with the community by publishing her queries, writing articles about her good practices or participating in conversations on technical issues. A data analyst can also seek knowledge from the community by viewing queries authored and published by other peers, searching for relevant articles or asking for help in the conversation board. By fostering collaboration and knowledge sharing, the platform facilitates organizational learning and innovation. It supports diverse activities, including expertise location, resource reuse, and social networking among analysts \cite{matthews2014goals,matthews2013community,muller2012diversity,rowe2012behaviour}. These features make it a representative example of modern collaborative technologies in large-scale industrial settings, enabling both novice and expert analysts to improve their productivity and skills.

\subsection{Data and Measures}
\subsubsection{Dataset}
The raw dataset for this study comprises two primary sources:

\begin{enumerate} 
\item \textbf{Usage Data:} Backend logs automatically collected from a collaborative data platform, capturing analysts’ interactions over a four-year period from January 2014 to March 2018. All user identifiers were hashed to ensure privacy. The dataset includes: 
\begin{itemize} 
\item \textbf{Query Records:} Metadata for all queries, such as query ID, title, author (hashed user ID), time of creation, brief descriptions, and publication status. 
\item \textbf{Execution Records:} Logs of query executions, detailing query ID, hashed user ID, execution timestamps, and query length. 
\item \textbf{Viewing Records:} Comprehensive records of user interactions, documenting views of query pages on the digital platform. 
\end{itemize} 
\item \textbf{Employee Information:} Organizational data available through the public company hub, including hashed user ID and the corresponding tenure, title, and hierarchical position. No additional personal identifiers were accessed or used in the study.
\end{enumerate}

The final dataset was curated with the following criteria:
\begin{itemize}
    \item \textbf{Inclusion Criteria}: Analysts who authored at least one query on the platform during the study period were included.
    \item \textbf{Exclusion Criteria}: We then excluded 1) platform administrators and admin employees of the platform vendor to avoid skewing the analysis with non-representative user behaviors; 2) Analysts with missing organizational data (e.g., due to employment termination during the study period) were also removed.
\end{itemize}
After applying the inclusion and exclusion criteria, the dataset initially comprised 2,059 analysts. These analysts were categorized by organizational levels based on their reporting structure, ranging from entry-level employees to senior staff. To ensure representativeness, extremely senior or junior analysts were excluded, resulting in a final sample of 2,027 active analysts who collectively authored 101,327 queries during the study period. Queries that were never executed or contained incomplete metadata were subsequently removed, leaving a final dataset of 79,797 queries written by 2,001 data analysts. Notably, only about one in eight queries was ever viewed by an analyst other than the original author, with 10,049 (approximately 1 in 8) queries authored by 1,097 analysts being reused by peers.

\subsubsection{Variables}
\paragraph{Dependent variable}
To develop a productivity measure for data analysts in our study, we borrow the concept of \emph{Pre-alpha} phase from the software development life cycle \cite{tiwari2010some,piggot2013healthy,buse2008metric}.The development process consists of several phases, as summarized in prior research \cite{rothfuss2002framework,tiwari2010some}. During the \emph{Pre-alpha} phase, developers focus on writing preliminary source code, a critical step that takes place before Alpha testing. 
We define \emph{FirstCompletionTime$_{i, k}$} as the time interval between the point when the data analyst $i$ clicks the button to create an empty query $k$ and the point when she executes this query for the first time:
\begin{equation}
\text{FirstCompletionTime}_\text{analyst $i$, query $k$}
= \text{Timestamp}_\text{$i$ first executes}^{k} -
\text{Timestamp}_\text{$i$ creates}^{k}
\end{equation}
This time interval, comparable to the \emph{Pre-alpha} phase in software development, characterizes the time that a data analyst spends to shape her idea into a testable, prototype query. The longer such time interval is, the later this analyst can proceed the following steps.  We, therefore, use \emph{FirstCompletionTime$_{i, k}$} as a proxy for data analyst productivity. These query writing start time-stamps and the first execution time-stamps are automatically logged at the back end of the collaborative data platform. We believe that the data analysts in our study cannot manipulate this data directly nor have no incentive to act strategically, as only the administrators of the collaborative data platform are informed and have access to these data.

We also include other variables related with analyst's participation in learning activities (i.e. Learning activities$_{it}$ and variables related with analyst's productivity in writing query (i.e. $Z_{itk}$ in Equation \ref{eq-ch2-NBmean}). 

\paragraph{Variables Related with Learning Activities}
\begin{itemize}
	\item \emph{Number of Queries Written$_{it}$}: this is defined as the number of queries that a data analyst $i$ has written by herself on the collaborative data platform during period $t$. The mean of \emph{Number of Queries Written$_{it}$} is $0.782$ and the standard deviation is $3.82$.
   \item  \emph{Number of Queries Viewed$_{it}$}: this is defined as the number of queries that a data analyst $i$ has viewed from her peers on the collaborative data platform during period $t$. The mean of \emph{Number of Queries Viewed$_{it}$} is $0.306$ and the standard deviation is $2.03$.
\end{itemize}

\paragraph{Variables Related with Productivity in Writing Queries}
Prior work suggests that individual adeptness, multi-tasking and task complexity are likely to affect productivity. A more adept data analyst may write a query faster; a data analyst who has piles of work may become less productive because of stress and pressure; a data analyst may spend more time on writing a complex query that consists of many statements. We incorporate the variables to calculate the probabilities of state-dependent productivity in writing queries. We explain the definitions of these control variables for the scenario in which the analyst $i$ creates and first executes query $k$.
\begin{itemize}
\item \emph{Workload$_{i,k}$:} This is the average number of queries that analyst $i$ composes simultaneously together with the focal query $k$ during the \emph{FirstCompletionTime$_{i,k}$}. This definition is inspired by the workload developed in \cite{tan2014does} for restaurant workers. 
For example, suppose \emph{FirstCompletionTime$_{i, k}$} lasts 40 minutes. During this period, the author data analyst only creates another query that overlaps with the focal query $k$ for 20 minutes. The \emph{Workload$_{i, k}$}, therefore, is (40 min + 20 min)/(40 min) = 1.5 queries.
\item \emph{Query Size$_{k}$:} This is the number of statements that were executed in the first execution of query $k$. 
\item \emph{Saved Query$_{k}$:} This is a binary variable that indicates whether query $k$ has been saved. 
\item \emph{Migrated Query$_{k}$:} This is a binary variable that indicates whether part of query $k$ was migrated from a different platform. We acquire such information by parsing the title or description of query $k$.
\item \emph{Tenure on collaborative data platform$_{i, k}$:} This is the number of months between the date when the author analyst $i$ joined the collaborative data platform and the time she creates query $k$.
\end{itemize}

The descriptive statistics for these variables can be found in Table \ref{table:des_stat}. In addition to these variables, we include a state-specific constant in $Z_{itk}$. This constant term captures a state-specific fixed effect in query-writing productivity, which can be interpreted as 
the intrinsic productivity in writing queries for an average analyst in the corresponding state. 
\begin{center}{ 
 {\renewcommand{\arraystretch}{0.8}
\begin{longtable}{@{\extracolsep{5pt}}L{4cm}@{}c@{}c@{}c@{}c@{}c@{}c@{}} 
\caption{Descriptive statistics of the raw data capturing $N=79797$ queries written and executed by 2001 data analysts at the organization during 2014-2018.} 
  \label{table:des_stat} 
 \\
\toprule 
Statistic & \multicolumn{1}{c}{Mean} & \multicolumn{1}{c}{St. Dev.} & \multicolumn{1}{c}{Min} & \multicolumn{1}{c}{Pctl(25)} & \multicolumn{1}{c}{Pctl(75)} & \multicolumn{1}{c}{Max} \\ 
\midrule
\endfirsthead

\multicolumn{7}{c}%
{{\tablename\ \thetable{} -- continued from previous page}} \\
\midrule
Statistic & {Mean} & {St. Dev.} & {Min} & {Pctl(25)} & {Pctl(75)} & {Max} \\ 
\midrule
\endhead

\midrule \multicolumn{7}{r}{{Continued on next page}} \\ \bottomrule
\endfoot

\bottomrule
\endlastfoot

FirstCompletionTime & 90,711 & 1,200,935 & 1 & 20 & 262 & 83,798,568 \\ 
Aggregate Direct Experience & 116.2 & 200.2 & 0 & 16 & 126 & 1,775 \\ 
Aggregate Indirect Experience & 35.9 & 70.8 & 0 & 1 & 35 & 1,510 \\ 
Direct Experience with Different Databases & 52.4 & 124.8 & 0 & 2 & 49 & 1,389 \\ 
Direct Experience with the focal Database & 63.9 & 113 & 0 & 6 & 68 & 1,092 \\ 
Workload & 1.1 & 0.7 & 1 & 1 & 1 & 28 \\ 
Tenure on collaborative data platform & 14.9 & 13.1 & 0 & 4 & 24 & 50 \\
Saved Query & 0.3 & 0.5 & 0 & 0 & 1 & 1 \\ 
Migrated Query & 0.01 & 0.1 & 0 & 0 & 0 & 1 \\ 
Query Size & 4.6 & 117.6 & 1 & 1 & 1 & 18,095 \\
\end{longtable}}}
\end{center}

\subsection{Hidden Markov Model for Data Analysts Learning}
\subsubsection{Theory}
The latent nature and structure of such dynamic learnings make them very difficult to capture. To analyze the learning dynamics of data analysts in our setting, we propose a Hidden Markov Model (HMM) similar to that in \cite{singh2011hidden}. The HMM is a model of bivariate Markov process $\{S_t, O_t\}_{t=1}^{\infty}$ where $\{S_t\}_{t=1}^{\infty}$ is a Markov process that consists of a finite \emph{state} space and $\{O_t\}_{t=1}^{\infty}$ is a sequence of observations that is determined by a state-specific stochastic process. Generally $\{S_t\}_{t=1}^{\infty}$ is assumed not directly observable (\emph{hidden}). The possibility of modeling complex structures of $\{O_t\}_{t=1}^{\infty}$ through a simple formulation makes HMMs attractive and widely applied in areas ranging from bioinformatics \cite{leroux1992maximum}, ecology \cite{morf1998stochastic} and criminology, to speech recognition \cite{rabiner1986introduction} and transportation \cite{macdonald1997hidden, scott2002bayesian}. They are also found to be very important in econometrics and macro-economics. For example, \cite{hamilton1989new} proposed an HMM as a very tractable approach to modeling unobservable regime shifts and to estimating the impact of such shifts on the level of U.S. real gross national product. Another well-known tool in the quantitative finance literature with essentially the same structure of HMM is the Kalman filter \cite{koopman1997exact, duan1999estimating}.

  The HMM model described in this section is an individual-level model of query writing and learning behaviors. In our HMM, we characterize these \emph{hidden states} with respect to the latent productivity and learning ability of data analysts, or \emph{absorptive ability} in \cite{cohen1990absorptive}'s term. Such latent characteristics can include programming skills, knowledge on the databases and other factors that may influence analyst productivity and subsequent learning. At any point of time, a data analyst resides in only one state. The transitions between states over a period are affected by her participation in \emph{learning activities}. Such activities can include the data analyst's writing her own queries (i.e. \emph{learning by doing}) and her viewing on queries written by other analysts (i.e. \emph{learning by viewing}). The relationship of a state with a unique productivity behavior in writing query, along with the Markovian transitions between states allow us to capture data analyst's productivity and learning dynamics.
  
  To do so, we use a nonhomogeneous HMM \cite{hughes1999non} in which the Markovian transitions are a function of time-varying variables. This is important for us to understand the drivers of the dynamics rather than merely building a model that fits the dynamics in the data. Specifically, we first assume that a data analyst’s probability for moving from one state to another is determined by her observed participation in learning activities as well as the unobserved individual random effects. Second, the observation in our HMM is the productivity of individual analyst in writing a query in different periods of time. The query-specific productivity is a function of analyst's intrinsic productivity, observed characteristics of the query, and unobserved individual random effects. Adding random effects to both parts of the model grants us the possibility to capture the heterogeneity of analysts' performance in both learning and writing queries. 
 \begin{figure}[tb]
 \centering
    \includegraphics[width=\linewidth]{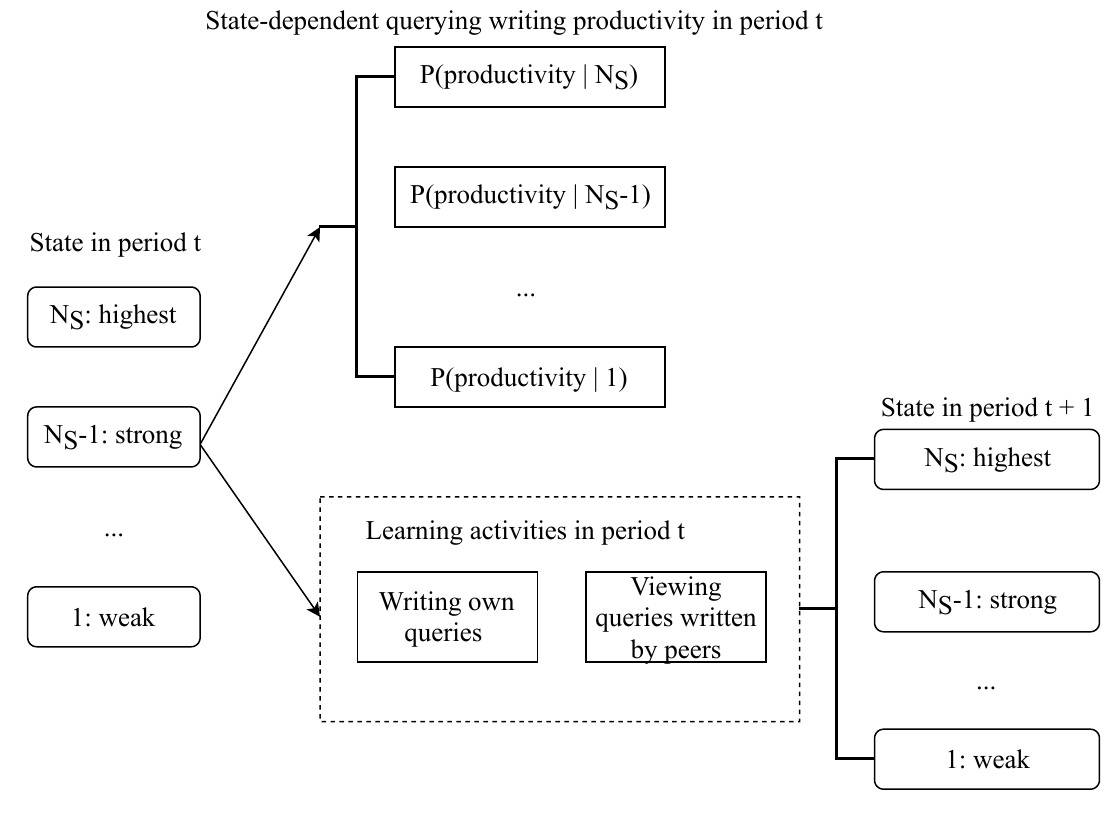}
    \caption{A Hidden Markov Model of Data Analyst Learning and Writing Query}
    \label{fig:ch2_hmmtimeline}
     
\end{figure}

Figure \ref{fig:ch2_hmmtimeline} shows how our HMM relates the transitions between latent states to observed dynamics in productivity. Specifically, in period $t$, an analyst probabilistically resides in one of total $N_s$ states based on her accumulated learning stock. Each realized state corresponds to a unique stochastic process that characterizes the analyst's productivity in writing queries in period $t$. Her participation in learning activities during period t determines whether she will move up to a higher state, stay where she is, or move down to a lower state in the next period.

This model has several advantages compared with the standard learning curve model. First, it allows us to capture the learning dynamics of a data analyst by investigating the differential impact of learning activities on her transitions between different hidden states, conditional on the state she currently resides. Second, this model also provides us a dynamic segmentation of data analysts based on their time-varying states, whereas an ad-hoc specification of is usually added to an otherwise static model. Third, the use of state and the inherent dependency between serial states within the model enable us to fully account the effect of autocorrelation. Autocorrelation in our setting may occur when an unobservable shock (e.g. unexpected illness of the individual analyst) commonly affects serial or multiple query-writing observations. This unobservable shock, if only exists in one period, can be captured by the state-specific query-writing process that is common to all observations in that period. If the shock spans across two serial periods, the state that corresponds to the previous period can carry on the effect of the unobservable shock to the current state thanks to the Markovian dependency between serial states. Fourth, in this model the current query-writing behavior only depends on the lag participation in learning activities. By doing so it diminishes reverse causality between learning and change in productivity. An analyst may view many queries from peers because she needs to write an extremely complex query. Such reverse causality may lead to erroneous estimate of learning effects in the classical static model.
\subsubsection{Model Development}
To analyze the learning dynamics and productivity of data analysts, we developed a Hidden Markov Model (HMM). This framework is designed to capture the evolution of an analyst’s latent learning states over time and link these hidden states to their observable query-writing productivity. In essence, the model enables us to infer the unseen “learning journey” of data analysts and connect it to measurable outcomes.

In our HMM, the latent states represent distinct levels of productivity and learning ability, ranging from novices (state 1) to highly proficient analysts (state $N_s$). These states are ordered to reflect progression as analysts accumulate skills and knowledge over time. The entire time horizon is divided into a finite number of periods, $t \in \{1, 2, ..., T\}$. At the beginning of any given period, a data analyst resides in a latent state. This state determines her productivity level for that period, and her participation in learning activities influences her transition to another state (or staying in the same state) in the next period.
For each data analyst $i$' we observe a sequence of query-writing productivity outcomes over time, denoted 
as $O^{i} = \{O^{i}_1, O^{i}_2, ..., O^{i}_T\}$, where $O^{i}_t$ represents the observed productivity in period $t$. These observed outcomes are assumed to be influenced by the hidden states and transitions are estimated. Below, we provide an overview of the key components of modeling framework, with detailed technical specifications available in the appendix.

\paragraph{The State Transition Probabilities}  
The transitions between latent learning states are modeled as a Markov process where only transitions to adjacent states are feasible. Specifically, the conditional probability that analyst \(i\) moves from state \(s\) in period \(t\) to state \(s'\) in period \(t+1\) is denoted as \(q^{it}_{s, s^{'}} = P(S^i_{t+1} = s' | S^i_{t} = s)\). This structure reflects the gradual nature of learning progression and regression, ensuring that analysts cannot “skip” states in their learning journey. 

Transitions between states are determined by the \emph{stock of learning} accumulated by the analyst during the period. The stock of learning, denoted as \(\mathit{LS}^{it}_{s}\), represents the cumulative effect of participation in learning activities, which include both \emph{learning by doing} (e.g., writing queries) and \emph{learning by observing} (e.g., reviewing queries authored by others). The following equation defines how the stock of learning determines the likelihood of state transitions, emphasizing the role of learning activities and unobserved heterogeneity:
\[
\mathit{LS}^{it}_{s} = \matr{\beta}_s \text{Learning activities}_{it} + \zeta_i + e_{ist},
\]
where \(\text{Learning activities}_{it}\) is a vector capturing analyst \(i\)'s participation in various learning activities during period \(t\). The vector \(\matr{\beta}_s\) represents state-specific coefficients that quantify the effect of these activities on building the learning stock. This state-specific design reflects the intuition that analysts in higher states may derive diminishing returns from the same learning activities compared to those in lower states, as suggested by research on program cognition and learning effects \cite{von1995program}. 

The term \(\zeta_i\) represents analyst-specific random effects, capturing unobserved heterogeneity such as participation in offline learning activities or differences in baseline abilities. The residual term \(e_{ist}\) accounts for random fluctuations in learning stock and is assumed to follow an extreme value distribution. 

The transition probabilities are modeled using an \emph{ordered logit framework}\cite{greene2010modeling}, where the likelihood of moving to a higher or lower state depends on whether the learning stock exceeds a state-specific threshold. Analysts move up to a higher state if their learning stock during the period exceeds a threshold \(\mu(s+1, s)\), move down to a lower state if their learning stock falls below \(\mu(s-1, s)\), or remain in their current state otherwise. These thresholds are ordered to ensure consistency:
\[
\mu(s+1, s) > \mu(s-1, s), \quad \forall s \in \{1, 2, ..., N_s\}.
\]
Additionally, boundary constraints are imposed to ensure that analysts in the highest or lowest states transition appropriately (e.g., \(\mu(N_s+1, N_s) = \infty\) and \(\mu(0, 1) = -\infty\)).

This formulation provides a behaviorally interpretable framework for understanding the drivers of learning dynamics. By linking participation in learning activities to transitions between states, the model highlights how targeted efforts (e.g., engaging in peer query reviews or creating more queries) influence analysts' progression or regression in latent learning states. Furthermore, the ordered logit structure ensures that transitions remain probabilistically coherent while capturing the nuanced effects of learning activities and individual differences.

\paragraph{State-Dependent Productivity in Writing Queries}  
We use \emph{FirstCompletionTime$_{k}^{it}$} as a proxy for data analyst productivity. Specifically, it is the time interval between the point when the data analyst $i$ clicks the button to create an empty query $k$ and the point when she executes this query for the first time, in period $t$::
\[
\text{FirstCompletionTime}_{k}^{it}
= \text{Timestamp}_\text{$i$ first executes $k$} -
\text{Timestamp}_\text{$i$ creates the empty $k$}.
\]
 Given that \emph{FirstCompletionTime$_{k}^{it}$} is a count variable and suffers from over-dispersion, i.e. the variance is significantly larger than the mean, we assume it follows a state-dependent negative binomial distribution:
\[
P(\text{FirstCompletionTime$_{k}^{it}$} = \tau| S^i_t = s) = \frac{e^{-\lambda_{itk}}\lambda_{itk}^{\tau}}{\Gamma(\tau + 1)},
\]
where \(\tau\) represents the observed completion time, and \(\lambda_{itk}\) is the conditional mean, defined as:
\[
\log(\lambda_{itk}) = \matr{\rho}_{s} Z_{itk} + \eta_i + \epsilon_{itk}.
\]

Here, \(Z_{itk}\) is a vector of observable, time-varying factors that influence analyst \(i\)'s productivity in writing query \(k\) during period \(t\). The coefficients \(\matr{\rho}_s\) are state-specific, capturing how these external factors impact productivity differently depending on the analyst’s current latent state \(s\). For instance, analysts in higher latent states may exhibit greater resilience to external challenges due to accumulated learning and experience.

The term \(\eta_i\) represents analyst-specific random effects, encapsulating unobserved heterogeneity such as individual programming style or domain expertise \cite{blikstein2014programming}. Meanwhile, \(\epsilon_{itk}\) is a log-gamma distributed error term, capturing residual variation not explained by \(Z_{itk}\) or \(\eta_i\).

To accommodate the over-dispersed nature of the data, we reformulate the probability distribution using a state-specific dispersion parameter \(\delta_s > 0\), the log-gamma distribution of the error term $\epsilon_{itk}$ \cite{greene2011econometric}
\[
P(\text{FirstCompletionTime$_{k}^{it}$} = \tau| S^i_t = s) = \frac{\Gamma(\delta_s + \tau)}{\Gamma(\tau + 1)\Gamma(\delta_s)} h_{itk}^{\delta_s} (1 - h_{itk})^{\tau},
\]
where
\[
h_{itk} = \frac{\delta_s}{e^{\rho_{s}Z_{itk} + \eta_i} + \delta_s}.
\]

\paragraph{The Likelihood of an Observed Sequence of Productivity Outcome}
Note that in our empirical setting, a data analyst can write multiple queries in one period. That is, given a data analyst $i$ has written $K$ queries in period $t$, the productivity outcome we can observe in the period is a sequence of the \emph{FirstCompletionTime} as the follows:
 \begin{equation}
 	O^i_t = \{\text{FirstCompletionTime$_{1}^{it}$}, \text{FirstCompletionTime$_{2}^{it}$}, ..., \text{FirstCompletionTime$_{K}^{it}$}\}
 \end{equation}
  Therefore, the probability of observing $O_t^i$ is:
 
 \begin{equation}
 	P(O^i_t | S^i_t = s) = \prod_{k = 1,2,...,K}P(\text{FirstCompletionTime$_{k}^{it}$}| S^i_t = s)
 \end{equation} 
  Because of the Markovian structure of this model, the probabilities of an individual's productivity outcome are correlated through the underlying sequence of the hidden states. Specifically, consider the realized state path for an analyst $i$ is $S^i = \{S_1^i, S_2^i, ..., S_T^i\}$ and the observed sequence of outcome $O^i = \{O_1^i, O_2^i, ..., O_T^i\}$. Then we have the likelihood of observing the sequence of $O^i$ is
\begin{equation}
  \label{eq-ch2-likelihood-sum}
  	L(O^i) = \sum_{S^i_1 = 1}^{n}\sum_{S^i_2 = 1}^{n}...\sum_{S^i_T = 1}^{n}P(S^i, O^i)
  \end{equation}
The full likelihood across analysts is simply the production of individual likelihood over $i \in \{1,2,...,N_{\text{ind}}\}$. 

\subsubsection{Estimation Strategy}  
Two main approaches have been proposed for estimating the parameters of Hidden Markov Models (HMMs): (1) maximum likelihood estimation, typically achieved through direct optimization or the Expectation-Maximization (EM) algorithm \cite{dempster1977maximum, baum1972inequality}, and (2) Bayesian estimation. For this study, we employ a hierarchical Bayesian estimation procedure, as described in \cite{rossi2012bayesian}, due to its flexibility in capturing cross-sectional heterogeneity among data analysts.

Bayesian estimation offers several key advantages for modeling learning dynamics and productivity. By integrating prior information with observed data, it provides robust inference and generates posterior distributions of model parameters, enabling a richer characterization of variability. This is particularly valuable for analyzing individual-level differences in state transitions and productivity patterns, as Bayesian methods account for heterogeneity more effectively than traditional approaches. Prior research \cite{heckman1981heterogeneity} emphasizes the risks of neglecting such heterogeneity, which can lead to spurious or overstated effects in dynamic models.

To implement Bayesian estimation, we use Markov Chain Monte Carlo (MCMC) techniques, including Gibbs sampling and the Metropolis-Hastings algorithm \cite{metropolis1953equation, hastings1970monte, 4767596}. These methods efficiently explore the posterior distribution, even for models with complex dependencies and random effects. Detailed descriptions of the MCMC algorithms and their implementation are provided in the appendix, including the rationale for their selection in this study. Briefly, Gibbs sampling is employed for parameters with tractable conditional distributions, while Metropolis-Hastings complements this by handling parameters without closed-form solutions. Together, these techniques ensure accurate and computationally efficient estimation of both state-specific and individual-level effects.

\section{Results}
\begin{table}[tb]
\sisetup{table-format=7.1, round-mode=places, round-precision=1, table-number-alignment=center}
\centering
\begin{threeparttable}
\caption{Fit Measures for Comparing Models}
\label{tb:ch2-model-comp}
{\renewcommand{\arraystretch}{1}
\begin{tabularx}{0.8\textwidth}{@{}lSSS@{}}
\toprule
 &{-2LogLikelihood} & {AIC} &{BIC}\\
\midrule
 One-state Model &1232380.8 &1232400.8 &1232493.6 \\
 Three-state Model &433477.1166181923 & 441551.1166181923&479043.70929716097\\
\bottomrule
\end{tabularx}
\begin{tablenotes}
      \small
      \item Note: The more negative the number, the better the fit. AIC, Akaike information criterion; BIC, Bayesian information criterion.    \end{tablenotes}
}
\end{threeparttable}
\end{table}

  We estimate two models of a single state and of three states respectively. The one state model assumes contemporaneous effect of query writing and viewing on analyst productivity. It is equivalent to the classic learning curve model. We use the maximum likelihood procedure to estimate this model. 
  In contrast, the three-state model accounts for learning dynamics over time and was estimated using a hierarchical Bayesian approach, implemented via Markov Chain Monte Carlo (MCMC) methods. Specifically, the Gibbs sampler and random-walk Metropolis-Hastings algorithm were employed, leveraging the Julia programming language for computation \cite{Julia-2017}. The MCMC procedure was initialized with random values, running for 100,000 iterations across two chains until convergence was achieved. Convergence was assessed using the ratio of within-chain variance to between-chain variance for each parameter, following established guidelines \cite{brooks1998general}. The first 85,000 iterations were discarded as burn-in, and the remaining 15,000 iterations were retained for inference.
 
 Table \ref{tb:ch2-model-comp} compares the fit measures for both models, demonstrating that the three-state model consistently outperforms the single-state model across all criteria. This underscores the importance of incorporating the latent learning states of analysts to better capture the complexity of their productivity and learning behaviors. The results suggest that modeling learning as a dynamic process provides a more nuanced and accurate understanding of the interaction between query writing, query viewing, and productivity improvements. 
  
  \subsection{HMM Estimates}
  Table \ref{tb:ch2-HMM-1} presents the summary statistics, including means, standard deviation (Std) and $95\%$ highest posterior density (HPD) intervals, calculated from the posterior distribution for all parameters. Estimates with HPDs that do not overlap zero are considered statistically significant.

\begin{table}[!htbp]
\sisetup{
  table-format=3.3,       % Ensures numbers have 3 decimal places
  round-mode=places,
  round-precision=3,      % Round to 3 decimal places
  table-number-alignment=center,
  detect-weight=true
}
\centering
\begin{threeparttable}
\caption{Estimation Results for the Three-state Hidden Markov Models Parameter}
\label{tb:ch2-HMM-1}
{\renewcommand{\arraystretch}{1}
\begin{tabularx}{0.9\textwidth}{@{}l
                                 S[table-format=3.3]  % Posterior mean
                                 S[table-format=3.3]  % Posterior SD
                                 l                    % Parenthesized intervals as text
                                 l@{}}
\toprule
Parameter & {Posterior mean} & \multicolumn{1}{c}{Posterior Std} & \multicolumn{2}{c}{$95\%$ HPD intervals} \\
\midrule
$\beta_1^{\text{writing}}$ & \bfseries 5.864 & 1.416 & (4.450, & 8.351) \\
$\beta_2^{\text{writing}}$ & \bfseries 4.379 & 1.763 & (2.542, & 7.537) \\
$\beta_3^{\text{writing}}$ & 0.338 & 0.377 & (-0.127, & 0.988) \\
$\beta_1^{\text{viewing}}$ & \bfseries 5.267 & 1.163 & (4.021, & 7.001) \\
$\beta_2^{\text{viewing}}$ & \bfseries -8.368 & 0.814 & (-9.964, & -7.480) \\
$\beta_3^{\text{viewing}}$ & \bfseries -8.306 & 1.440 & (-10.411, & -6.482) \\
$\mu(2,1)$ & \bfseries 20.547 & 1.585 & (17.167, & 22.590) \\
$\mu(2,3)$ & -3.458 & 2.906 & (-6.975, & 0.383) \\
$\mu(1,2)$ & \bfseries -13.262 & 2.006 & (-16.626, & -11.211) \\
$\text{log}(\mu(3,2) - \mu(1,2))$ & \bfseries 12.955 & 0.784 & (11.431, & 13.996) \\
$\text{log}\delta_1$ & \bfseries \num{-0.000180} & \num{0.000175} & (\num{-0.000667}, & \num{-0.000005}) \\
$\text{log}\delta_2$ & \bfseries 10.759 & 0.806 & (9.759, & 12.003) \\
$\text{log}\delta_3$ & \bfseries -2.287 & 0.186 & (-2.782, & -2.081) \\
$\rho_1^{\text{constant}}$ & \bfseries 5.847 & 2.351 & (3.667, & 10.039) \\
$\rho_2^{\text{constant}}$ & \bfseries 3.218 & 0.577 & (2.462, & 4.019) \\
$\rho_3^{\text{constant}}$ & \bfseries -6.842 & 0.636 & (-7.844, & -6.183) \\
$\rho_1^{\text{workload}}$ & \bfseries 9.884 & 0.530 & (8.925, & 10.788) \\
$\rho_2^{\text{workload}}$ & 2.794 & 2.304 & (-0.072, & 6.611) \\
$\rho_3^{\text{workload}}$ & \bfseries 7.655 & 1.212 & (5.347, & 8.734) \\
$\rho_1^{\text{tenure of author}}$ & \num{0.006} & 0.057 & (-0.087, & 0.117) \\
$\rho_2^{\text{tenure of author}}$ & \bfseries -8.147 & 1.756 & (-11.554, & -6.130) \\
$\rho_3^{\text{tenure of author}}$ & \bfseries -8.975 & 1.780 & (-11.075, & -6.648) \\
$\rho_1^{\text{migrated query}}$ & 2.318 & 3.203 & (-1.430, & 6.844) \\
$\rho_2^{\text{migrated query}}$ & \bfseries -4.642 & 1.830 & (-6.962, & -2.277) \\
$\rho_3^{\text{migrated query}}$ & \bfseries 9.185 & 0.434 & (8.305, & 9.676) \\
$\rho_1^{\text{saved query}}$ & \bfseries 0.283 & 0.094 & (0.177, & 0.544) \\
$\rho_2^{\text{saved query}}$ & \bfseries 8.404 & 2.942 & (4.814, & 12.674) \\
$\rho_3^{\text{saved query}}$ & \num{-0.011} & 0.644 & (-1.031, & 1.186) \\
$\rho_1^{\text{query size}}$ & -3.518 & 2.373 & (-8.646, & 0.002) \\
$\rho_2^{\text{query size}}$ & -4.815 & 2.519 & (-9.414, & 0.629) \\
$\rho_3^{\text{query size}}$ & 0.217 & 0.329 & (-0.092, & 0.960) \\
\bottomrule
\end{tabularx}
\begin{tablenotes}
      \small
      \item Note: Bolded coefficients’ HPDs do not cross zero.
    \end{tablenotes}
}
\end{threeparttable}
\end{table}

 \subsubsection{State Dependent Productivity}
  Our analysis reveals clear variations in state-specific intrinsic productivity among analysts, as indicated by the constant terms $\rho_1^{\text{constant}}, \rho_2^{\text{constant}},$ and $\rho_3^{\text{constant}}$. The posterior means for these constants are $5.8$, $3.2$, and $-6.8$ for states 1, 2, and 3, respectively, all of which are statistically significant. These results suggest that intrinsic productivity improves as analysts transition to higher states (a more negative constant implies shorter \emph{FirstCompletionTime}). Based on this pattern, we interpret the states as follows: analysts in state 1 exhibit characteristics of inexperience and slower query-writing productivity, which we label as \emph{novice}. Similarly, analysts in state 2, with moderate productivity, are categorized as \emph{intermediate}, while state 3, reflecting the highest productivity, is labeled as \emph{advanced}.

  \paragraph{Workload Effects}
 The posterior means for workload effects are $9.9$, $2.8$, and $7.7$ for states 1, 2, and 3, respectively. These positive values indicate that higher workloads decrease analysts' productivity across all states. However, the effect is smallest in magnitude and statistically insignificant for analysts in the intermediate state. This suggests that analysts at the intermediate level may better manage workloads compared to their novice or advanced counterparts.

 \paragraph{Tenure Effects}
  
The posterior means for the effect of analysts' tenure on the platform are $0.006$, $-8.1$, and $9.0$ for states 1, 2, and 3, respectively. These results indicate that tenure positively impacts productivity for analysts in the intermediate and advanced states. However, the effect is insignificant for analysts in the novice state, suggesting that tenure benefits may not manifest until analysts gain sufficient experience or skills.

\paragraph{Query-Specific Characteristics}

Our results highlight the relationship between certain query-specific characteristics and \emph{FirstCompletionTime}. One notable factor is whether a query is \textit{migrated} (originating on another platform and later moved to the current system). Our results show that analysts in novice state are insensitive to whether the current query is migrated or not. In contrast, intermediate-state analysts exhibit significant productivity gains on migrated queries, with a significant posterior mean of $-4.6$. Conversely, analysts in the advanced state show a significant slowdown when working on migrated queries. One potential explanation for this pattern is analysts in advanced state are usually asked to migrate large and complex queries. Ensuring those queries work appropriately on a new platform requires more time and caution as more calibration and comments may be necessary. 

Similarly, our results show that queries saved by analysts (potentially signaling complexity) increase the time required to complete the query for novice and intermediate analysts. However, this effect is insignificant for advanced analysts, suggesting that they are less hindered by query complexity. Interestingly, we observe no significant effect of query size on \emph{FirstCompletionTime} across all states. This lack of effect may stem from the fact that most queries (more than $75\%$) in our study consist of only one statement, which likely minimizes variability in query size effects.

  \subsubsection{Learning and State Transitions}
The estimated thresholds for state transitions provide valuable insights into the learning dynamics of data analysts. Transitioning from the novice to the intermediate state requires reaching a threshold of $20.5$, while moving down from the advanced to the intermediate state occurs at a threshold of $-3.5$. For analysts in the intermediate state, the thresholds are $-13.3$ to regress to the novice state and an exceptionally high $422,932.8$ to advance to the advanced state. These thresholds represent the accumulation of learning activities, such as writing or viewing queries, needed for an analyst to transition between states. The results show that progressing to higher states becomes increasingly difficult as analysts advance in their learning. Specifically, moving from the intermediate to the advanced state requires far greater effort compared to progressing from the novice to the intermediate state.

  \paragraph{Effects of Learning Activities}
The impact of learning activities on state transitions reveals nuanced, state-dependent effects. For novice analysts, both writing their own queries and viewing queries authored by peers significantly increase the likelihood of transitioning to the intermediate state. However, for intermediate analysts, while the effect of writing queries remains positive, its magnitude diminishes. This observation aligns with prior research, such as \cite{shafer2001effects}, which highlights that individuals with higher expertise often experience slower rates of subsequent learning within their area of specialization.

In contrast, analysts in the intermediate state or higher exhibit an increased likelihood of transitioning to a lower state when they engage in viewing peers' queries. This counterintuitive pattern may stem from the variability in the quality of published queries. Here we only consider the aggregate views for an analyst without dif-
ferentiating the source of knowledge. Analysts may encounter poorly designed or irrelevant queries that detract from their productivity. Moreover, viewing peer queries, though seemingly beneficial, often competes with the time available for actual programming or query-writing tasks, creating delays and inefficiencies in productive work.

  \paragraph{Cognitive and Behavioral Implications}
This competition for time and attention is particularly pronounced for analysts in higher states, who may derive diminishing returns from additional peer-based learning. These findings resonate with the behavioral observation that "a wealth of information creates a poverty of attention" \cite{anderson1997situative}. Analysts in advanced states may find the abundance of available materials less helpful because of the cognitive burden associated with evaluating and integrating new information. For these individuals, the effort required to sift through peers’ queries may outweigh the benefits, especially when the materials are not immediately relevant or aligned with their expertise.

Another plausible explanation for this pattern is domain switching. When intermediate or advanced analysts engage more frequently in viewing peers’ queries, it may indicate a shift in their responsibilities or focus, such as transitioning to a new business domain or database. This change often necessitates a temporary return to novice-like performance as analysts acclimate to unfamiliar contexts. In such cases, increased viewing activity reflects the challenges of adapting to new information and workflows rather than a decline in skill. Consequently, these observations emphasize the importance of considering the contextual and temporal aspects of learning behaviors when interpreting productivity dynamics.

  \section{Discussion}
Classical literature on individual learning curves often assumes a static model, which risks producing biased estimates of learning effects when individual behavior evolves dynamically. This paper presents a Hidden Markov Model (HMM) to capture these dynamics in the productivity and learning behaviors of data analysts. By comparing our model to the standard learning curve framework, we demonstrate that the HMM is a more appropriate approach for our empirical setting. The HMM effectively captures three ordered learning states through which analysts progress, revealing increasing intrinsic productivity as they transition to higher states.

Our findings underscore the importance of participation in two distinct learning activities—\emph{writing queries (learning from direct experience)} and \emph{viewing peers' queries (learning from indirect experience)}—in shaping analysts' transitions between these states. The results indicate that the effects of these activities differ depending on the current learning state of the analyst, providing valuable insights into the nuanced nature of learning dynamics.

\subsection{Design Implications for Systems and Interfaces}
This study makes several significant contributions to the design and improvement of data platforms and collaborative systems, particularly in optimizing user productivity and learning trajectories. By leveraging insights from the proposed Hidden Markov Model (HMM), we underscore the importance of mechanisms for identifying user states, guiding learning activities, and delivering personalized experiences.
\begin{itemize}
    \item \textbf{Mechanisms to Identify User Learning States}: Our framework demonstrates the feasibility of detecting an analyst’s latent learning state based on their productivity and activity patterns. Integrating real-time state detection into platforms enables targeted support tailored to each user’s needs. For instance, novice analysts can receive prompts for viewing peer queries and guided tutorials, while advanced analysts might benefit from features that streamline the creation of original, complex queries. These mechanisms lay the foundation for adaptive interfaces that evolve with the user’s learning progress.
    \item \textbf{Guided Learning Activities for Optimized Productivity} The differential effects of learning activities across states underscore the need for systems to prioritize state-appropriate activities. Platforms should dynamically recommend activities aligned with the user’s current state. For novice analysts, a mix of writing and viewing queries could enhance foundational skills, while intermediate users may benefit from features that encourage independent problem-solving. For advanced analysts, tools that minimize cognitive distractions—such as personalized filters or curated peer query recommendations—can help sustain high productivity.
    \item \textbf{Personalized Mechanisms for Learning Progression}: Platforms can leverage predictive capabilities of the model to anticipate when analysts are likely to transition between states. For example, analysts approaching a state transition could receive motivational feedback, skill-building tasks, or curated learning resources aligned with their progression trajectory. Additionally, platforms can identify behaviors, such as excessive peer query viewing, that may hinder growth for advanced users. Visualizing progress through dashboards that display state-specific milestones and learning goals can further engage users, providing both a sense of accomplishment and actionable guidance.
    \item \textbf{Cost-Effective User Development and Onboarding}: The insights from this study can inform the design of onboarding programs that accelerate analysts’ transitions from novice to advanced states. Tailored learning paths, curated based on an analyst’s initial state and progression, can reduce the time and resources required to achieve proficiency. Automating aspects of learning support also minimizes the need for costly external training, allowing organizations to invest in scalable, efficient development solutions.
    \item \textbf{Enhancing Collaboration without Overload}: The observed risks of cognitive overload for analysts in higher states highlight the importance of managing information flow effectively. Features such as intelligent query recommendations, relevance-based filtering, and quality indicators for peer contributions can mitigate the negative effects of excessive peer query viewing. By prioritizing high-value collaborative activities and reducing irrelevant distractions, platforms can ensure that users derive maximum benefit from peer-based learning while maintaining their productivity..
\end{itemize}
By implementing these mechanisms, platforms can foster an environment that not only optimizes individual productivity but also enhances the overall learning experience. These contributions underscore the role of adaptive and personalized design in maximizing the potential of technology-mediated work and collaboration.
  
\subsection{Limitations and Future Directions}
This study provides significant insights into the dynamic learning and productivity of data analysts through the application of a Hidden Markov Model (HMM). However, there are several limitations that warrant further exploration. First, we restricted our analysis to one-state and three-state HMMs. While our results indicate that a dynamic model outperforms static models, the three-state configuration may not fully capture the complexity of analyst behavior. Future work should explore HMMs with varying numbers of states to determine the optimal model configuration that best represents the data. Second, our assumption of a uniform initial state distribution may be overly simplistic. Incorporating a more flexible approach by estimating the initial state probabilities at the population level could enhance the model's accuracy and robustness, as suggested in prior work \cite{lee2017personal}.
Additionally, our hierarchical modeling framework could be extended to incorporate individual random effects as functions of observable characteristics, such as demographics, organizational roles, or geographic locations \cite{allenby1995using, netzer2008hidden}. This extension would allow us to explore how specific individual or contextual factors influence learning dynamics and state transitions, offering a more nuanced understanding of variability across users. Furthermore, while this study focuses on aggregate learning activities, future research could refine the analysis by examining the quality and relevance of viewed queries, particularly in advanced states, where cognitive overload risks are prominent.

Lastly, the generalizability of our findings should be validated across other collaborative systems, organizational settings, and broader categories of knowledge workers. While this study focuses on data analysts, many other knowledge workers—such as software engineers, research scientists, or technical writers—may share similar learning behaviors, including the interplay of independent work and collaborative learning. Extending the framework to examine these groups would provide valuable insights into how learning dynamics and productivity manifest in diverse professional contexts. Moreover, testing the model across different platforms and industries could uncover platform-specific or organizational influences on learning and productivity. For instance, variations in collaborative tools, team structures, or organizational cultures might alter the effectiveness of learning activities and the observed state transitions. Testing the framework across such settings would not only validate its robustness but also refine its applicability.

By addressing these limitations and pursuing these directions, future work can deepen our understanding of knowledge workers' learning dynamics while providing actionable insights for the design of more effective collaborative and adaptive systems.

\section{Disclosure}
This research did not receive any specific grant and no potential competing interest was reported by the author(s). 

\section{Data Availability Statement}
Data supporting the findings of this study are not publicly available due to confidentiality agreements and non-disclosure obligations with the data provider. As such, the dataset cannot be shared.

\bibliographystyle{tfq}
\bibliography{interacttfqsample}

\appendix
\section{HMM Model Development}
Consider a finite set of states $s \in \{1, 2, ..., N_s\}$ in our HMM, with 1 being the lowest state and $N_s$ being the highest. The entire time horizon is divided in to a finite number of periods $t \in \{1, 2, ..., T\}$. At the beginning of any given period, a data analyst resides in an unknown state. For each period and each data analyst, we observe the productivity of the data analyst in writing queries, other factors that can affect her productivity and her participation in various learning activities during the period. The observed sequence of a data analyst $i$'s productivity over periods is denoted as her outcome sequence $O^{i} = \{O^{i}_1, O^{i}_2, ..., O^{i}_T\}$, where $O^{i}_t$ characterizes her temporal productivity in writing queries in period $t$. Given these primitives, our HMM has three main components:
    \begin{enumerate}
        \item The initial state distribution, $\matr{\pi}$;
        \item The sequence of the state-transition probability matrix, $\matr{Q}$, with $\matr{Q}_{t-1 \rightarrow t}$ denoting the Markovian transition matrix from $t-1$ to $t$;
        \item The observed productivity with their state-dependent probability ${P}$. The probability that data analyst $i$ is observed with certain productivity at time $t$ conditional on her state is $P(O_t^i|S_t^i)$, where $S_t^i$ is the state that $i$ belongs to in period t and $O_t^i$ is the observed productivity of $i$ in the same period.
    \end{enumerate}	
    
    \subsection{The Initial State Distribution}
    The initial state distribution characterizes the probability that an average data analyst starts in a particular state at the beginning of time horizon (period $t = 1$). Let $\pi_s$ be the probability that an average analyst belongs to the state $s$ in period $t = 1$, where $s \in \{1, 2, ..., n\}$. We have $\sum_{s = 1, 2, ..., N_s}\pi_s = 1$. In our HMM, we assume initially the probabilistic membership of all data analysts to different states follows the same distribution:
    \begin{equation}
    	\matr{\pi} = (\pi_1, \pi_2, ... \pi_{N_s})
    \end{equation}
    Many studies assuming time-homogeneous HMM define the initial state distribution as the stationary distribution of the transition matrix \cite{macdonald1997hidden, netzer2008hidden, montoya2010climate}. However, in our HMM the transition matrix is time-varying, thereby violating the assumption of homogeneity. Some other studies estimate their HMMs with fixed initial state distribution. Such assumption can come from prior information \cite{schweidel2011portfolio} or is simply specified for the computational convenience \cite{singh2011hidden}. An alternative shown in a recent study by \cite{lee2017personal} is to directly estimate the initial state probability distribution at the population level. In our HMM we fix the initial state distribution to be uniform across states, i.e. $\pi_s = \frac{1}{N_s}, \forall s \in \{1, 2, ..., N_s\}$.
    \subsection{The State Transition Matrix}
     We model the transitions between latent learning states as a Markov process where only transitions to adjacent states are feasible. The transition matrix is defined as the follows:
     \begin{equation}
     	\matr{Q}^i_{t \rightarrow t+1} = \begin{array}{c@{\!\!\!}l}
  \overbrace{\left( \begin{array}[c]{cccccc}
    q^{it}_{1,1} & q^{it}_{1,2} & 0 & \cdots & 0 &0 \\
    q^{it}_{2,1} & q^{it}_{2,2} & q^{it}_{2,3} & \cdots & 0 & 0\\
    \vdots & \vdots & \vdots & & \vdots & \vdots \\
    0 & 0 & 0 & \cdots & q^{it}_{N_s,N_s-1}& q^{it}_{N_s,N_s} \\
  \end{array}  \right)}^{\text{\normalsize States:$1, 2, ..., N_s$}}  
&
 \begin{array}[c]{@{}l@{\,}l}
\left. \begin{array}{c} 
\vphantom{\begin{array}{c}1\\1\\1\\1\end{array}}
\end{array} \right\} & \text{States:} 1, 2, ..., N_s
\end{array}
\end{array}
     \end{equation}
 where $q^{it}_{s, s^{'}} = P(S^i_{t+1} = s' | S^i_{t} = s)$ is the conditional probability that analyst $i$ moves from state $s$ in period $t$ to state $s'$ in period $t$, abd $0 \leq q^{it}_{s, s^{'}} \leq 1 \quad \forall s, s^{'}, \quad \sum_{s^{'} = 1, 2, ..., N_s} q^{it}_{s, s^{'}} = 1 \quad \forall s$. We model the transitions between the states as a threshold model; what transition occurs depends on how the stock of learning compares to the threshold. 
The stock of learning, termed as $\mathit{LS}^i_{s}$ is assumed to be a function of analyst $i$'s participation in learning activities during the period as the follows:
 \begin{equation}
 	\mathit{LS}^{it}_{s} = \matr{\beta}_s \text{Learning activities}_{it} + \zeta_i + e_{ist}
 \end{equation} 
 where $\text{Learning activities}_{it}$ is vector of $i$'s participation in learning activities in this period and $\beta_s$ is a vector of coefficients capturing the effect of participation in these activities on building up the analyst's learning stock. Note that the effects of participation in learning activities are state-specific. Researchers in program cognition models have shown that past experiences and knowledge base of a programmer greatly affect her program understanding, thereby affecting the return of learning \cite{von1995program}. $\zeta_i$ is the random effect that captures the unobserved heterogeneity in state transitions across analysts. Such unobserved heterogeneity may include participation in offline learning activities that we do not observe. Assuming the unobserved $e_{ist}$ is independently and identically distributed (IID) of the extreme value type, the non-homogeneous transition probabilities are modeled as an ordered logit \cite{greene2010modeling}:
 
 \begin{equation}
\begin{split}
    q^{it}_{s, s+1} = &1 - \frac{exp(\mu(s+1,s) - \matr{\beta}_s \text{Learning activities}_{it} - \zeta_i)}{1 + exp(\mu(s+1,s) - \matr{\beta}_s \text{Learning activities}_{it} - \zeta_i}\\
    q^{it}_{s, s-1}  = &\frac{exp(\mu(s-1,s) - \matr{\beta}_s \text{Learning activities}_{it} - \zeta_i)}{ 1 + exp(\mu(s-1,s) - \matr{\beta}_s \text{Learning activities}_{it} - \zeta_i)}\\
    q^{it}_{s, s} = &\frac{exp(\mu(s+1,s) - \matr{\beta}_s \text{Learning activities}_{it} - \zeta_i)}{1 + exp(\mu(s+1,s) - \matr{\beta}_s \text{Learning activities}_{it} - \zeta_i} \\
    &- \frac{exp(\mu(s-1,s) - \matr{\beta}_s \text{Learning activities}_{it} - \zeta_i)}{ 1 + exp(\mu(s-1,s) - \matr{\beta}_s \text{Learning activities}_{it} - \zeta_i)}
    \end{split}
     \end{equation}
  where $\mu(s+1,s)$ is the logit threshold for analysts in state $s$ to move up to state $s+1$ and $\mu(s-1,s)$ the logit threshold for analysts in state $s$ to move down to state $s-1$. Put it into words, a data analyst will move up to a higher state if her stock of learning during the period is greater than $\mu(s+1,s)$, or move down to a lower state if her temporal learning is smaller than $\mu(s-1,s)$. We constrain $\mu(s+1,s) > \mu(s-1,s), \forall s \in \{1, 2, ..., n\}$ to ensure the order of these thresholds. We also fix $\mu(N_s+1,n) = \infty$ and $\mu(0, 1) = - \infty$ to ensure that analysts already in the lowest or highest state transit correctly. Analysts of states $s\in \{2, 3, ..., N_s-1\}$ can move up to a higher state, move down to a lower state or stay in the current state.
  \subsection{State-Dependent Productivity in Writing Queries}
  We use \emph{FirstCompletionTime$_{k}^{it}$} as a proxy for data analyst productivity. For the detailed definition of \emph{FirstCompletionTime$_{k}^{it}$}, see 1.3.3.1. Specifically, it is the time interval between the point when the data analyst $i$ clicks the button to create an empty query $k$ and the point when she executes this query for the first time, in period $t$:
\begin{equation}
\text{FirstCompletionTime}_{k}^{it}
= \text{Timestamp}_\text{$i$ first executes $k$} -
\text{Timestamp}_\text{$i$ creates the empty $k$}
\end{equation}
Because  \emph{FirstCompletionTime$_{k}^{it}$}  is a count variable and suffers from over-dispersion, i.e. the variance is significantly larger than the mean, we assume it follows a state-dependent negative binomial distribution:
\begin{equation}
\label{eq-ch2-NB1}
	P(\text{FirstCompletionTime$_{k}^{it}$} = \tau| S^i_t = s) = \frac{e^{-\lambda_{itk}}\lambda_{itk}^{\tau}}{\Gamma(\tau + 1)}
\end{equation}
  where the conditional mean $\lambda_{itk}$ is defined as the following:
  \begin{equation}
  \label{eq-ch2-NBmean}
  	log(\lambda_{itk}) = \matr{\rho}_{s}Z_{itk} + \eta_i + \epsilon_{itk}
  \end{equation}
   where $Z_{itk}$ is a vector of time-varying factors that can affect the productivity of analyst $i$ in writing query $k$ in period $t$. $\matr{\rho}_s$ is a vector of state-specific coefficients that capture the effect of $Z_{itk}$, where $s$ is the state that $i$ resides in period $t$. The random effect $\eta_i$ captures the analyst-specific unobserved heterogeneity, which may include individual programming style \cite{blikstein2014programming}. With the log-gamma distribution of the error term $\epsilon_{itk}$, we can reformulate equation \ref{eq-ch2-NB1} as the following \cite{greene2011econometric}:
   \begin{equation}
   	P(\text{FirstCompletionTime$_{k}^{it}$} = \tau| S^i_t = s) = \frac{\Gamma(\delta_s + \tau)}{\Gamma(\tau + 1)\Gamma(\delta_s)}h_{itk}^{\delta_s}(1 - h_{itk})^{\tau}
   \end{equation}
  where
  \begin{equation}
  	h_{itk} = \frac{\delta_s}{e^{\rho_{s}Z_{itk} + \eta_i} + \delta_s}
  \end{equation}
  with $\delta_s > 0$ as a state-specific dispersion parameter.
 
 Note that in our empirical setting, a data analyst can write multiple queries in one period. That is, given a data analyst $i$ has written $K$ queries in period $t$, the productivity outcome we can observe in the period is a sequence of the \emph{FirstCompletionTime} as the follows:
 \begin{equation}
 	O^i_t = \{\text{FirstCompletionTime$_{1}^{it}$}, \text{FirstCompletionTime$_{2}^{it}$}, ..., \text{FirstCompletionTime$_{K}^{it}$}\}
 \end{equation}
  Therefore, the probability of observing $O_t^i$ is:
 
 \begin{equation}
 	P(O^i_t | S^i_t = s) = \prod_{k = 1,2,...,K}P(\text{FirstCompletionTime$_{k}^{it}$}| S^i_t = s)
 \end{equation}
 
  \subsection{The Likelihood of an Observed Sequence of Productivity Outcome}
  Because of the Markovian structure of this model, the probabilities of an individual's productivity outcome are correlated through the underlying sequence of the hidden states. Specifically, consider the realized state path for an analyst $i$ is $S^i = \{S_1^i, S_2^i, ..., S_T^i\}$ and the observed sequence of outcome $O^i = \{O_1^i, O_2^i, ..., O_T^i\}$. Then we have
  \begin{equation}
  	P(O^i|S^i) = \prod_{t = 1}^{T}P(O^i_t|S^i_t)
  \end{equation}
  And the probability of seeing such state path is:
  \begin{equation}
  	P(S^i) = P(S^i_1)\prod_{t = 1}^{T-1}q^{it}_{S^i_t, S^i_{t+1}}
  	\end{equation}
  	where $P(S^i_1)$ is the probability that the initial state of $i$ is $S^i_1$; here $P(S^i_1 = s) = \pi_s$. So the joint probability of $S^i$ and $O^i$ is
  	\begin{equation}
  	\label{eq-ch2-singlepath}
  		P(S^i, O^i) = P(S^i_1)\prod_{t = 1}^{T-1}q^{it}_{S^i_t, S^i_{t+1}}\prod_{t = 1}^{T}P(O^i_t|S^i_t)
  	\end{equation}
  Accordingly, the likelihood of observing the sequence of $O^i$ is given by summing equation \ref{eq-ch2-singlepath} over all possible state paths that $i$ could take overtime:
  \begin{equation}
  \label{eq-ch2-likelihood-sum}
  	L(O^i) = \sum_{S^i_1 = 1}^{n}\sum_{S^i_2 = 1}^{n}...\sum_{S^i_T = 1}^{n}P(S^i, O^i)
  \end{equation}
  Following \cite{macdonald1997hidden}, we can simplify equation \ref{eq-ch2-likelihood-sum} into the following matrix notation:
  \begin{equation}
  	L(O^i) = \matr{\pi} \matr{P^i_1}\matr
  	{Q^i_{1\rightarrow2}}\matr{P^i_2}\matr
  	{Q^i_{2\rightarrow3}}\cdots\matr{P^i_T}\matr{1^{'}}
  \end{equation}
  where 
  \begin{equation}
  \label{eq-ch2-indlikelihood}
  \matr{P^i_t}= \begin{array}{c@{\!\!\!}l}
  {\left( \begin{array}[c]{cccccc}
    P(O^i_t | S^i_t = 1) & 0 & 0 & \cdots & 0 &0 \\
    0 & P(O^i_t | S^i_t = 2) & 0 & \cdots & 0 & 0\\
    \vdots & \vdots & \vdots & & \vdots & \vdots \\
    0 & 0 & 0 & \cdots & 0& P(O^i_t | S^i_t = N_s) \\
  \end{array}  \right)}  
\end{array}
  \end{equation}
  and $\matr{1^{'}}$ is an $N_s \times 1$ vector of ones.
  The full likelihood across analysts is simply the production of individual likelihood over $i \in \{1,2,...,N_{\text{ind}}\}$.

\section{HMM Estimation Strategy} 

We estimate our HMM using a standard hierarchical Bayes estimation procedure described in \cite{rossi2012bayesian}, for its great flexibility and efficiency in estimating models that account for cross-sectional heterogeneity. 
Instead of focusing on the point estimate of true parameter value as the frequentist approach, the main interest of the Bayes approach is in generating the entire distribution of the parameters given the data and a prior. Bayes theorem provides the mechanism for how data and prior beliefs are translated in to posterior beliefs: 
\begin{equation}
\label{eq-ch2-bayes}
	p(\theta|y) = \frac{p(y|\theta)p(\theta)}{p(y)} \propto p(y|\theta)p(\theta)
\end{equation}
where $\theta$ is the parameters we want to estimate and $p(\theta|y)$ is the posterior distribution of $\theta$, given observed data $y$ and prior $p(\theta)$. Inference here refers to summarizing the posterior distribution with typical statistics, such as the posterior mean $E[\theta] = \int\theta p(\theta|y) \it{d}\theta$ and posterior standard deviation. Note that bayesian inference here is conducted only using formal probability theory and does not require asymptotic approximations, which is particularly useful in modeling data with limited size. 
    
    To apply a standard hierarchical Bayes estimation procedure, we distinguish two sets of parameters: random-effect parameters $\{\mathbf{\Theta_i}\}$ and parameters that are common across individuals $\mathbf{\Psi}$. Specifically,
    \begin{equation}
    \begin{split}
    	\matr{\Theta_i} =& \{\zeta_i, \eta_i\}\\
    	\matr{\Psi} =& \{\mu(2,1), \mu(1,2), \mu(3,2), ..., \mu(s-1,s), \mu(s+1,s), ..., \mu(N_s-1,N_s),\\
    	     & \delta_1, \delta_2, ..., \delta_s,..., \delta_{N_s},\\\
    	     & \matr{\beta_1}, \matr{\beta_2}, ..., \matr{\beta_s}, ..., \matr{\beta}_{N_s},\\
    	     & \matr{\rho}_1, \matr{\rho}_2, ..., \matr{\rho}_s,...,\matr{\rho}_{N_s}\}
    \end{split}
    \end{equation}
   and we assume a prior distribution for the random parameters as the following:
 \begin{equation}
        \text{Prior}(\matr{\Theta_i}) = \text{Normal} (\matr{0}, \matr{\Sigma_{\Theta}})
                \end{equation}
Conforming to convention, we complete this specification by assuming the typical diffuse priors for $\Psi$ and the hyperparameter $\matr{\Sigma}_{\theta}$, following \cite{netzer2008hidden}
\begin{equation}
	\begin{split}
		\text{Prior}(\matr{\Sigma_{\Theta}}) =& \text{Inverse Wishart}(N_{\Theta} + 5 + N_{ind}, \matr{I_{N_{\Theta}}})\\
		\text{Prior}(\matr{\Psi}) = & \text{Normal}(\matr{0}, \matr{30I_{N_{\Psi}}})
	\end{split}
\end{equation}
Conducting bayes estimation requires sequential drawing from the conditional posterior $p(\{\matr{\Theta_i}\}, \matr{\Psi}|y)$ using the Markov Chain Monte Carlo (MCMC) algorithm. Note that Equation \ref{eq-ch2-bayes} along with Equation \ref{eq-ch2-indlikelihood} suggests that the conditional posterior does not have a closed form. Thus, we apply Metropolis-Hasting algorithm \cite{metropolis1953equation, hastings1970monte} to sample from the posterior distribution, using a Gibbs sampler \cite{4767596}.

The basis of the Metropolis algorithm consists of: (1) in the $m^{th}$ iteration, simulate a candidate sample of parameters, say $\Omega^{cand}$, from the proposal distribution $\Phi^{m-1}(\cdot)$ based on the last accepted sample $\Omega^{m-1}$; (2) accept the candidate sample with the probability calculated via the acceptance function $\alpha(\Omega^{cand}|\Omega^{m-1})$. We choose the Gaussian distributions as the proposal distributions, taking advantage of its symmetry. That is $\Phi^{m-1}(\cdot) = \text{Normal}(\Omega^{m-1}, \Sigma)$. The typical acceptance function is the following:
\begin{equation}
	\alpha(\Omega^{cand}|\Omega^{m-1}) = \text{min}\{1, \frac{P(\Omega^{cand}|y)}{P(\Omega^{m-1}|y)}\}
\end{equation}
where $P(\cdot|y)$ is the posterior joint density given the data. This acceptance function is designed to send the sampler to move to higher probability areas under the posterior joint density. In practice, we draw a random number uniformly between 0 and 1, and compare it with $\frac{P(\Omega^{cand}|y)}{P(\Omega^{m-1}|y)}$. If $\frac{P(\Omega^{cand}|y)}{P(\Omega^{m-1}|y)}$ is larger than the uniform random number, we accept the candidate sample; otherwise we reject it.  After running the algorithm for a long time, we end up with a sequential sample from the target posterior distribution of the parameters. 

Gibbs sampler is an MCMC sampler developed by \cite{4767596}, which allows us to divide variables of interest into blocks consisting of two or more variables and sample the joint distribution of each block conditional on all other variables. Gibbs sampler is particularly suitable for hierarchical models. With Gibbs sampler, we can recursively update the individual random effects $\{\matr{\Theta_i}\}$, the covariance $\matr{\Sigma_{\theta}}$, and the common effects $\matr{\Psi}$ following Algorithm \ref{ch2-hmm-bayes-algo}:

\begin{algorithm}
\DontPrintSemicolon
\SetKwInOut{Init}{Initialize}
\Init{Generate a random sample of \\
$\matr{\Sigma_{\Theta}}^{0} \sim \text{Inverse Wishart}(N_{\Theta} + 5 + N_{ind}, \matr{I_{N_{\theta}}})$\\
 $\{\matr{\Theta_i}^{0}\} \sim \text{Normal} \left( \matr{0}, \matr{\Sigma_{\Theta}}^{0} \right)$\\
 $\matr{\Psi}^{0} \sim  \text{Normal}(\matr{0}, \matr{30I_{N_{\Psi}}})$}
 \For{$\text{iteration } m = 1, 2, \cdots$}{
 \nlset{Step 1} \For{$i \leftarrow 1 $ \KwTo $N_{\text{ind}}$}{
  \nl Simulate a candidate $\matr{\Theta_i}^{\text{cand}} \sim \text{Normal}(\matr{\Theta_i^{m-1}}, \lambda_{\theta} \widetilde{\matr{\Sigma}_{\theta}})$, where $\matr{\Theta_i^{m-1}}$ is the accepted draw in iteration $m-1$, and $\lambda_{\theta}$ and $\widetilde{\matr{\Sigma}_{\theta}}$ are chosen adaptively following the Algorithm 4 in \cite{andrieu2008tutorial} and \cite{shaby2010exploring}. We adapt $\lambda_{\theta}$ and $\widetilde{\Sigma_{\theta}}$ to reduce the autocorrelation between draws and to maintain the optimal acceptance rate.\;
  \nl Accept $\matr{\Theta_i}^{\text{cand}}$ to be $\matr{\Theta_i}^{m}$ with the probability of 
  \[\text{min}\left \{\frac{\text{exp}(-\frac{1}{2}\matr{\Theta_i}^{\text{cand}}(\matr{\Sigma_{\Theta}}^{m-1})^{-1}\matr{\Theta_i}^{'\text{cand}})P_{i}(y_i|\matr{\Theta_i}^{\text{cand}}, \matr{\Psi}^{m-1})}{\text{exp}(-\frac{1}{2}\matr{\Theta_i}^{m-1}(\matr{\Sigma_{\Theta}}^{m-1})^{-1}\matr{\Theta_i}^{'m-1})P_{i}(y_i|\matr{\Theta_i}^{m-1}, \matr{\Psi}^{m-1})}, 1\right \}\]
  if reject, $\matr{\Theta_i}^{m} = \matr{\Theta_i}^{m-1}$.
 }
 \nlset{Step 2} Sample $\matr{\Sigma_{\Theta}}^{m} \sim \text{Inverse Wishart}(N_{\Theta} + 5 + N_{ind}, \matr{I_{N_{\Theta}}} + \sum_1^{N_{\text{ind}}} \matr{\Theta_{i}^{m}}\matr{\Theta_{i}^{'m}})$\;
 \nlset{Step 3} Given the updated $\{\matr{\Theta_i}^{m}\}$\;
    \nlset{1} Simulate a candidate $\matr{\Psi}^{\text{cand}} \sim \text{Normal}(\matr{\Psi^{m-1}}, \lambda_{\Psi} \widetilde{\matr{\Sigma_{\Psi}}})$, where $\matr{\Psi^{m-1}}$ is  the accepted draw in iteration $m-1$. $\lambda_{\Psi}$ and $\widetilde{\matr{\Sigma_{\Psi}}}$ are also adaptively chosen using the same algorithm as in Step 1. \;
 \nlset{2} Accept $\matr{\Psi}^{\text{cand}}$ as $\matr{\Psi}^{m}$ with the probability of
 
  \[\text{min}\left \{\frac{\text{exp}(-\frac{1}{2}\matr{\Psi}^{\text{cand}}(\matr{V_{\Psi}}^{0})^{-1}\matr{\Psi}^{'\text{cand}})P_{i}(y_i|\matr{\Theta_i}^{m}, \matr{\Psi}^{\text{cand}})}{\text{exp}(-\frac{1}{2}\matr{\Psi}^{m-1}(\matr{V_{\Psi}}^{0})^{-1}\matr{\Psi}^{'m-1})P_{i}(y_i|\matr{\Theta_i}^{m}, \matr{\Psi}^{m-1})}, 1\right \}\]
    where $\matr{V_{\Psi}}^{0} = \matr{30I_{N_{\Psi}}}$. If reject, $\matr{\Psi}^{m} = \matr{\Psi}^{m-1}$.
 
  }
\caption{Hierarchical Bayes Estimation Algorithm}
\label{ch2-hmm-bayes-algo}
\end{algorithm}

\end{document}